\documentclass{article}[10pt]

\usepackage{natbib}

\usepackage{eucal}
\usepackage{natbib}
\usepackage{amsfonts}
\usepackage{mathrsfs}
\usepackage{graphicx}
\usepackage{psfrag}
\usepackage{latexsym}
\usepackage{amsmath}
\usepackage{amstext}
\usepackage{amssymb}
\usepackage{array}
\usepackage{tabularx}
\usepackage{pifont}
\usepackage{psfrag}
\usepackage{color}
\begin{document}
%
\def\bcolon{\mbox{\boldmath$\colon$}}
\def\bdot{\mbox{\boldmath$\cdot$}}
\def\bone{{\bf 1}}
\def\bzero{{\bf 0}}
\def\DIV {\hbox{\rm div}}
\def\Grad{\hbox{\rm Grad}}
\def\sym{\mathop{\rm sym}\nolimits}
\def\dev{\mathop{\rm dev}\nolimits}
\def\Dev{\mathop {\rm DEV}\nolimits}
\def\jmpdelu{{\lbrack\!\lbrack \Delta u\rbrack\!\rbrack}}
\def\jmpudot{{\lbrack\!\lbrack\dot u\rbrack\!\rbrack}}
\def\jmpu{{\lbrack\!\lbrack u\rbrack\!\rbrack}}
\def\jmphi{{\lbrack\!\lbrack\varphi\rbrack\!\rbrack}}
\def\ljmp{{\lbrack\!\lbrack}}
\def\rjmp{{\rbrack\!\rbrack}}
\def\sB{{\mathcal B}}
\def\sU{{\mathcal U}}
\def\sC{{\mathcal C}}
\def\sS{{\mathcal S}}
\def\sV{{\mathcal V}}
\def\sL{{\mathcal L}}
\def\sO{{\mathcal O}}
\def\sH{{\mathcal H}}
\def\sG{{\mathcal G}}
\def\sM{{\mathcal M}}
\def\sN{{\mathcal N}}
\def\sF{{\mathcal F}}
\def\sW{{\mathcal W}}
\def\sT{{\mathcal T}}
\def\sR{{\mathcal R}}
\def\sJ{{\mathcal J}}
\def\sK{{\mathcal K}}
\def\sE{{\mathcal E}}
\def\sS{{\mathcal S}}
\def\sH{{\mathcal H}}
\def\sD{{\mathcal D}}
\def\sG{{\mathcal G}}
\def\sP{{\mathcal P}}
\def\Bnabla{\mbox{\boldmath$\nabla$}}
\def\BGamma{\mbox{\boldmath$\Gamma$}}
\def\BDelta{\mbox{\boldmath$\Delta$}}
\def\BTheta{\mbox{\boldmath$\Theta$}}
\def\BLambda{\mbox{\boldmath$\Lambda$}}
\def\BXi{\mbox{\boldmath$\Xi$}}
\def\BPi{\mbox{\boldmath$\Pi$}}
\def\BSigma{\mbox{\boldmath$\Sigma$}}
\def\BUpsilon{\mbox{\boldmath$\Upsilon$}}
\def\BPhi{\mbox{\boldmath$\Phi$}}
\def\BPsi{\mbox{\boldmath$\Psi$}}
\def\BOmega{\mbox{\boldmath$\Omega$}}
\def\Balpha{\mbox{\boldmath$\alpha$}}
\def\Bbeta{\mbox{\boldmath$\beta$}}
\def\Bgamma{\mbox{\boldmath$\gamma$}}
\def\Bdelta{\mbox{\boldmath$\delta$}}
\def\Bepsilon{\mbox{\boldmath$\epsilon$}}
\def\Bzeta{\mbox{\boldmath$\zeta$}}
\def\Beta{\mbox{\boldmath$\eta$}}
\def\Btheta{\mbox{\boldmath$\theta$}}
\def\Biota{\mbox{\boldmath$\iota$}}
\def\Bkappa{\mbox{\boldmath$\kappa$}}
\def\Blambda{\mbox{\boldmath$\lambda$}}
\def\Bmu{\mbox{\boldmath$\mu$}}
\def\Bnu{\mbox{\boldmath$\nu$}}
\def\Bxi{\mbox{\boldmath$\xi$}}
\def\Bpi{\mbox{\boldmath$\pi$}}
\def\Brho{\mbox{\boldmath$\rho$}}
\def\Bsigma{\mbox{\boldmath$\sigma$}}
\def\Btau{\mbox{\boldmath$\tau$}}
\def\Bupsilon{\mbox{\boldmath$\upsilon$}}
\def\Bphi{\mbox{\boldmath$\phi$}}
\def\Bchi{\mbox{\boldmath$\chi$}}
\def\Bpsi{\mbox{\boldmath$\psi$}}
\def\Bomega{\mbox{\boldmath$\omega$}}
\def\Bvarepsilon{\mbox{\boldmath$\varepsilon$}}
\def\Bvartheta{\mbox{\boldmath$\vartheta$}}
\def\Bvarpi{\mbox{\boldmath$\varpi$}}
\def\Bvarrho{\mbox{\boldmath$\varrho$}}
\def\Bvarsigma{\mbox{\boldmath$\varsigma$}}
\def\Bvarphi{\mbox{\boldmath$\varphi$}}
\def\bA{\mbox{\boldmath$ A$}}
\def\bB{\mbox{\boldmath$ B$}}
\def\bC{\mbox{\boldmath$ C$}}
\def\bD{\mbox{\boldmath$ D$}}
\def\bE{\mbox{\boldmath$ E$}}
\def\bF{\mbox{\boldmath$ F$}}
\def\bG{\mbox{\boldmath$ G$}}
\def\bH{\mbox{\boldmath$ H$}}
\def\bI{\mbox{\boldmath$ I$}}
\def\bJ{\mbox{\boldmath$ J$}}
\def\bK{\mbox{\boldmath$ K$}}
\def\bL{\mbox{\boldmath$ L$}}
\def\bM{\mbox{\boldmath$ M$}}
\def\bN{\mbox{\boldmath$ N$}}
\def\bO{\mbox{\boldmath$ O$}}
\def\bP{\mbox{\boldmath$ P$}}
\def\bQ{\mbox{\boldmath$ Q$}}
\def\bR{\mbox{\boldmath$ R$}}
\def\bS{\mbox{\boldmath$ S$}}
\def\bT{\mbox{\boldmath$ T$}}
\def\bU{\mbox{\boldmath$ U$}}
\def\bV{\mbox{\boldmath$ V$}}
\def\bW{\mbox{\boldmath$ W$}}
\def\bX{\mbox{\boldmath$ X$}}
\def\bY{\mbox{\boldmath$ Y$}}
\def\bZ{\mbox{\boldmath$ Z$}}
\def\ba{\mbox{\boldmath$ a$}}
\def\bb{\mbox{\boldmath$ b$}}
\def\bc{\mbox{\boldmath$ c$}}
\def\bd{\mbox{\boldmath$ d$}}
\def\be{\mbox{\boldmath$ e$}}
\def\bff{\mbox{\boldmath$ f$}}
\def\bg{\mbox{\boldmath$ g$}}
\def\bh{\mbox{\boldmath$ h$}}
\def\bi{\mbox{\boldmath$ i$}}
\def\bj{\mbox{\boldmath$ j$}}
\def\bk{\mbox{\boldmath$ k$}}
\def\bl{\mbox{\boldmath$ l$}}
\def\bm{\mbox{\boldmath$ m$}}
\def\bn{\mbox{\boldmath$ n$}}
\def\bo{\mbox{\boldmath$ o$}}
\def\bp{\mbox{\boldmath$ p$}}
\def\bq{\mbox{\boldmath$ q$}}
\def\br{\mbox{\boldmath$ r$}}
\def\bs{\mbox{\boldmath$ s$}}
\def\bt{\mbox{\boldmath$ t$}}
\def\bu{\mbox{\boldmath$ u$}}
\def\bv{\mbox{\boldmath$ v$}}
\def\bw{\mbox{\boldmath$ w$}}
\def\bx{\mbox{\boldmath$ x$}}
\def\by{\mbox{\boldmath$ y$}}
\def\bz{\mbox{\boldmath$ z$}}

\title{Biological remodelling: Stationary energy, configurational change, internal variables and dissipation}
\author{K. Garikipati\thanks{Corresponding author, assistant professor, Department of
Mechanical Engineering, and Program in Applied Physics. {\tt
krishna@umich.edu}}, J. E. Olberding\thanks{Graduate research
assistant, Department of Biomedical Engineering.}, E. M.
Arruda\thanks{Professor, Department of Mechanical Engineering, and
Program in Macromolecular Science and Engineering.}, K.
Grosh\thanks{Associate professor, Departments of Mechanical
Engineering, and Biomedical Engineering.}, \\
H. Narayanan\thanks{Graduate research assistant, Department of
Mechanical Engineering.}, S. Calve\thanks{Graduate research
assistant, Program in Macromolecular Science and Engineering.}\\
University of Michigan, Ann Arbor, \\Michigan 48109, USA}
\maketitle


\begin{abstract}
Remodelling is defined as an evolution of microstructure or
variations in the configuration of the underlying manifold. The
manner in which a biological tissue and its subsystems remodel
their structure is treated in a continuum mechanical setting.
While some examples of remodelling are conveniently modelled as
evolution of the reference configuration (Case I), others are more
suited to an internal variable description (Case II). In this
paper we explore the applicability of stationary energy states to
remodelled systems. A variational treatment is introduced by
assuming that stationary energy states are attained by changes in
microstructure via one of the two mechanisms---Cases I and II. An
example is presented to illustrate each case. The example
illustrating Case II is further studied in the context of the
thermodynamic dissipation inequality.
\end{abstract}

\section{Introduction and background}\label{sect1}

The development of a biological tissue and its subsystems consists
of the distinct processes of \emph{morphogenesis, growth and
remodelling}, a classification suggested by \citet{Taber:95}. For
preciseness of mathematical formulation we have previously defined
and treated growth as consisting of only addition or depletion of
mass through processes of transport and reaction, possibly coupled
with mechanics \citep{growthpaper}. We define remodelling as
\emph{microstructural changes within the biological structure at
constant mass}. While remodelling and growth occur simultaneously
and in a coupled fashion in biological tissues, they can be
treated as separate processes for modelling purposes. Furthermore,
in certain situations, addition and depletion remain in balance,
maintaining constant mass. This is referred to as
\emph{homeostasis}, during which remodelling can occur.

There is, in fact, experimental evidence for a strict separation
between remodelling and growth in soft tissue: (\romannumeral 1)
\citet{StopakHarris:82} described the orientation of collagen
fibrils due to the forces exerted on them by the fibroblasts
(tendon cells) in a collagen gel. Growth processes of resorption
of existing fibrils and production of new ones with the preferred
orientation were not reported in their paper, suggesting that it
was the mechanical action of fibroblasts alone that resulted in
fibril orientation. In this example the fibril orientation can be
viewed as the microstructural quantity undergoing an evolution in
the absence of growth. Fibril reorientation driven by stress and
occurring independently of growth was attained in our laboratory:
A collagen gel was formed in a dish with polyethylene supports
fixed to the base. When fibroblasts were added to the gel they
exerted traction, thereby aligning the collagen fibrils. The
orientation that was obtained corresponded to the axis defined by
the supports. (\romannumeral 2) Another instance of
microstructural change in collagen fibrils is their longitudinal
and lateral fusion. \citet{Birketal:95,Birketal:97} reported an
abrupt change in the length of embryonic chicken tendons from
$\approx 40\;\mu$m to $\approx 120\;\mu$m, on the $17^\mathrm{th}$
day after fertilization. In this case, while growth obviously was
taking place in the embryonic tendon, micrographs verified the
abrupt change to be due to longitudinal (end-to-end) fusion of
smaller fibrils. This time scale, over which fusion took place, is
much smaller than the time scale of growth. We therefore propose
that, in a mathematical treatment, it is appropriate to view this
remodelling process as taking place in the absence of growth.

In contrast to these examples in soft tissue stands the case of
bone, wherein the macroscopic process that is usually labelled
``remodelling'' takes place by resorption of existing collagen
fibrils and production of new fibrils with a preferred
orientation. In this case, therefore, the finer scale processes do
fit our definition of growth. For the sake of conceptual and
mathematical clarity we will ignore all processes that require
growth at any scales in this paper, and focus upon a continuum
mechanical treatment of remodelling.

The literature in biomechanics and mathematical biology has a
number of papers concerning ``remodelling''
\citep{CowinHegedus:76,Taber:95,HarriganHamilton:1993,Seliktaretal:00,TaberHumphrey:01,AmbrosiMollica:02}.
Almost universally, however, these papers treat mass and density
changes and the mechanics---characterized by internal
stress---that is associated with them. Therefore, by our
definition, they describe \emph{growth}. An exception is
\citet{HumphreyRajagopal:02}, in which the notion of a natural
configuration is introduced. It bears some similarities with the
treatment in Section \ref{sect2} of this paper.

Another treatment, by \citet{Driessenetal1:03}, is specific to the
evolution of fiber orientation within a tissue. However, there are
at least two fundamental shortcomings in their model:
(\romannumeral 1) It is unable to distinguish between cubic
orthotropy and isotropy, and (\romannumeral 2) it predicts that in
a tissue that is undeformed from its reference state, the fiber
orientations evolve until a tensorial variable representing their
distribution reaches a particular value. We discuss their model
and present a critique of both these features in Section
\ref{sect4.2}.

The following is the organization of this paper: The mathematical
treatment for cases that are best described by smooth
configurational changes on the underlying manifold (Case I
remodelling) is presented in Section \ref{sect2}. A
one-dimensional example is presented in Section \ref{sect3} to
illustrate this formulation. The local reorientation of collagen
fibrils, including our experiments and the treatment via internal
variables is discussed in Section \ref{sect4}. Thermodynamic
dissipation is discussed in Section \ref{sect4.1.2}. The paper
concludes with a brief discussion in Section \ref{sect7}.

\section{Case I remodelling: Microstructural changes that alter the reference configuration}\label{sect2}

\begin{figure}[ht]
\psfrag{A}{\small $\Omega_0$} \psfrag{B}{\small $\Omega^\ast$}
\psfrag{C}{\small $\Omega$} \psfrag{D}{\small $\Bvarphi$}
\psfrag{G}{\small $\Bvarphi^\ast$} \psfrag{E}{\small $\Bchi$}
\centering{\includegraphics[width=10cm]{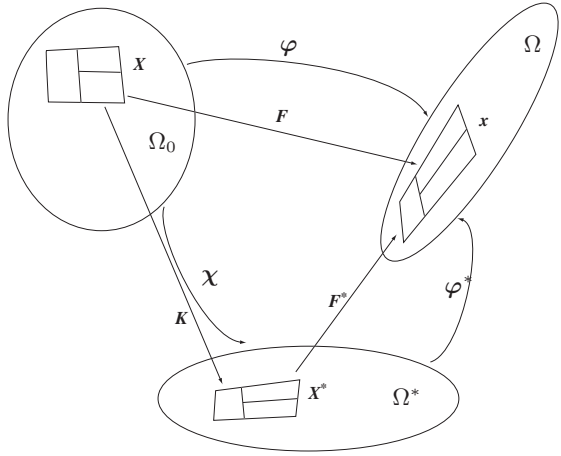}} \vskip -1cm
\caption{The kinematics of Case I remodelling and
deformation.}\label{fig3}
\end{figure}

Figure \ref{fig3} depicts the kinematics associated with Case I
remodelling. Material particles are labelled by $\bX \in
\mathbb{R}^3$ in the reference configuration, which is denoted by
$\Omega_0 \subset \mathbb{R}^3$. The material microstructure
undergoes changes that can be described by a point-to-point vector
map, $\Bchi\bcolon \Omega_0\times[0,T]\mapsto \mathbb{R}^3$,
defined as $\Bchi(\bX,t) \equiv \bX + \Bkappa(\bX,t)$. It carries
the microstructure from the reference configuration to a
remodelled configuration, $\Omega^\ast$, in which material points
will also be labelled as $\bX^\ast = \Bchi$. Assuming
$\Bchi(\bX,t)$ to be smooth in $\bX$, its tangent map is
$\bK(\bX,t) =
\partial\Bchi/\partial\bX$, leading to $\bK = \bone +
\partial\Bkappa/\partial\bX$. In Case I remodelling, therefore, $\bK$ denotes a
compatible change in configuration.\footnote{This configurational
change can be further decomposed, multiplicatively, into
incompatible components: $\bK = \bK^1\bK^2$ as in
\citet{remodelpaper}.}

Distinct from $\Bchi$ is the point-to-point vector map
$\Bvarphi^\ast\bcolon\Omega^\ast\times[0,T]\mapsto \mathbb{R}^3$.
It carries material points from $\Omega^\ast$ to the spatial
configuration $\Omega$, and is the deformation relative to
$\Omega^\ast$. The placement of material points in $\Omega$ is
therefore $\bx = \Bvarphi^\ast(\bX^\ast,t)$. The displacement,
$\bu^\ast$, satisfies $\Bvarphi^\ast(\bX^\ast,t) = \bX^\ast +
\bu^\ast(\bX^\ast,t)$. The classical deformation gradient is
$\bF^\ast :=
\partial\Bvarphi^\ast/\partial\bX^\ast = \bone +
\partial\bu^\ast/\partial\bX^\ast$. In this initial treatment we
do not consider any further decompositions of $\bF^\ast$. The
overall motion of a point is $\Bvarphi(\bX,t) = \Bkappa(\bX,t) +
\bu^\ast(\bX^\ast,t)\circ\Bchi(\bX,t)$, and the corresponding
tangent map is $\bF =
\partial\Bvarphi/\partial\bX$. It admits the multiplicative
decomposition $\bF = \bF^\ast\bK$.

\noindent{\bf Remark 1}: In applications, the microstructural
motion, $\Bchi$, will be distinguished from the deformation,
$\Bvarphi^\ast$, by either physical or mathematical
considerations. For instance, in the example of Section
\ref{sect3}, $\Bchi$ denotes the uncoiling of a long-chain
molecule, while $\Bvarphi^\ast$ is the stretching of bonds. A
mathematical decomposition can be motivated in other applications
such as $\Bchi$ denoting fine scale motion of material particles,
while $\Bvarphi^\ast$ denotes motion on coarser scales.

\subsection{A variational formulation}\label{sect2.1}

We wish to explore the relevance of stationary energy states to
remodelled biological systems. For this purpose the
microstructural changes of the type outlined with examples from
biology in Section \ref{sect1} and discussed more mathematically
in this section, are assumed to occur when the Gibbs free energy
of the system attains a stationary state. Importantly, the
stationary state of the biological system is not one of
equilibrium. The latter class of states is dictated by
thermodynamic dissipation \citep[see][]{DeGrootMazur:1984}. The
stationary energy states, on the other hand, can be identified via
standard variational arguments. For this purpose we consider the
following Gibbs free energy functional:
\begin{eqnarray}
\Pi_I[\bu^\ast,\Bkappa] &=&
\int\limits_{\Omega^\ast}\hat{\psi}^\ast(\bF^\ast,\bK,\bX^\ast)\mathrm{d}V^\ast\nonumber\\
& &
-\int\limits_{\Omega^\ast}\bff^\ast\bdot(\bu^\ast+\Bkappa)\mathrm{d}V^\ast
-\int\limits_{\partial\Omega^\ast_t}\bar{\bt}^\ast\bdot(\bu^\ast+\Bkappa)\mathrm{d}A^\ast,
\label{functnl}
\end{eqnarray}

\noindent where $\psi^\ast =
\hat{\psi}^\ast(\bF^\ast,\bK,\bX^\ast)$ is the Helmholtz free
energy density function, defined per unit volume in the remodelled
configuration. Observe that $\psi^\ast$ is assumed to depend upon
$\bK$ in addition to the usual dependence on $\bF^\ast$. Material
heterogeneity is allowed, and is represented by the dependence of
$\hat{\psi}^\ast$ on $\bX^\ast$. The body force per unit volume in
$\Omega^\ast$ is $\bff^\ast$, and the applied traction per unit
area of the surface subset, $\partial\Omega^\ast_t \subset
\partial\Omega^\ast$, is $\bar{\bt}^\ast$. We allow for displacement
boundary conditions, $\bu^\ast = \bg^\ast$, and vanishing
microstructural change, $\Bkappa = \bzero$ on
$\partial\Omega^\ast_u =
\partial\Omega^\ast\backslash\partial\Omega^\ast_t$. Since the total motion of a
material point is $\Bkappa + \bu^\ast$, the potential energy of
the external loads is as expressed by the second and third terms
in (\ref{functnl}).

\subsubsection{Euler-Lagrange equation: Quasistatic balance of linear
momentum}\label{sect2.1.1}

The functional (\ref{functnl}) yields one set of Euler-Lagrange
equations when stationarity of $\Pi_I$ is imposed with respect to
variations in $\bu^\ast$. We first define $\bu^\ast_\varepsilon :=
\bu^\ast + \varepsilon\delta\bu^\ast$, at fixed $\Bkappa$, with
$\delta\bu^\ast \in \mathbb{R}^3$ and $\delta\bu^\ast = \bzero$ on
$\partial\Omega^\ast_u$. Then,
\begin{eqnarray*}
\frac{\mathrm{d}}{\mathrm{d}\varepsilon}\Pi_I[\bu^\ast_\varepsilon,\Bkappa]
&=&
\frac{\mathrm{d}}{\mathrm{d}\varepsilon}\left(\int\limits_{\Omega^\ast}\hat{\psi}^\ast(\bF^\ast_\varepsilon,\bK,\bX^\ast)\mathrm{d}V^\ast - \int\limits_{\Omega^\ast}\bff^\ast\bdot(\bu^\ast_\varepsilon+\Bkappa)\mathrm{d}V^\ast\right)\\
& &
-\frac{\mathrm{d}}{\mathrm{d}\varepsilon}\int\limits_{\partial\Omega^\ast_t}\bar{\bt}^\ast\bdot(\bu^\ast_\varepsilon+\Bkappa)\mathrm{d}A^\ast.
\end{eqnarray*}
\noindent Differentiating under the integrals, applying the chain
rule, integrating by parts, and imposing stationarity via
$(\mathrm{d}\Pi_I/\mathrm{d}\varepsilon)_{\varepsilon=0}=0$ gives

\begin{eqnarray}
&&-\int\limits_{\Omega^\ast}\mathrm{Div}^\ast[\frac{\partial\psi^\ast}{\partial\bF^\ast}]\bdot\delta\bu^\ast\mathrm{d}V^\ast
-\int\limits_{\Omega^\ast}\bff^\ast\bdot\delta\bu^\ast\mathrm{d}V^\ast\nonumber\\
&&\qquad\qquad+\int\limits_{\partial\Omega^\ast_t}\left(\frac{\partial\psi^\ast}{\partial\bF^\ast}\bN^\ast\right)\bdot\delta\bu^\ast\mathrm{d}A^\ast
-\int\limits_{\partial\Omega^\ast_t}\bar{\bt}^\ast\bdot\delta\bu^\ast\mathrm{d}A^\ast
= 0.
\end{eqnarray}

\noindent Introducing $\bP^\ast =
\partial\psi^\ast/\partial\bF^\ast$, the arbitrariness of
$\delta\bu^\ast \in \mathbb{R}^3$ and the standard localization
argument yield the Euler-Lagrange equation
\begin{equation}
\mathrm{Div}^\ast\bP^\ast + \bff^\ast =
\bzero\;\mathrm{in}\;\Omega^\ast,\label{quasistatic}
\end{equation}

\noindent the following boundary condition and constitutive
relation
\begin{equation}
\bP^\ast\bN^\ast =
\bar{\bt}^\ast\;\mathrm{on}\;\partial\Omega^\ast_t,\quad\bP^\ast
\equiv\frac{\partial\psi^\ast}{\partial\bF^\ast}.\label{quasistatic1}
\end{equation}

\noindent Of these equations, (\ref{quasistatic}) is recognized as
the quasistatic balance of linear momentum in $\Omega^\ast$, and
(\ref{quasistatic1})$_1$ as the corresponding traction boundary
condition. Observe that, as defined in (\ref{quasistatic1})$_2$,
$\bP^\ast$ is the first Piola-Kirchhoff stress, which is conjugate
to $\bF^\ast$ with $\psi^\ast$ as the relevant strain energy
density function.

\subsubsection{Euler-Lagrange equation: Stationarity with respect
to microstructural change}\label{sect2.1.2}

The class of variations now considered is given by

\begin{equation}
\Bkappa_\varepsilon := \Bkappa + \varepsilon\delta\Bkappa,\quad
\mbox{at fixed}\; \bu^\ast, \;\mbox{with}\,\delta\Bkappa =
\bzero\;\mbox{on}\,\partial\Omega^\ast_u. \label{configvar}
\end{equation}

Proceeding as in Section \ref{sect2.1.1}:
\begin{eqnarray*}
\frac{\mathrm{d}}{\mathrm{d}\varepsilon}\Pi_I[\bu^\ast,\Bkappa_\varepsilon]
=
\frac{\mathrm{d}}{\mathrm{d}\varepsilon}\left(\int\limits_{\Omega^\ast}\hat{\psi}^\ast(\bF^\ast_\varepsilon,\bK_\varepsilon,\bX^\ast_\varepsilon)\mathrm{d}V^\ast - \int\limits_{\Omega^\ast}\bff^\ast\bdot(\bu^\ast+\Bkappa_\varepsilon)\mathrm{d}V^\ast\right)\qquad&&\\
 -
\frac{\mathrm{d}}{\mathrm{d}\varepsilon}\int\limits_{\partial\Omega^\ast_t}\bar{\bt}^\ast\bdot(\bu^\ast+\Bkappa_\varepsilon)\mathrm{d}A^\ast,&&\label{varint}
\end{eqnarray*}

\noindent where the following relations hold:
\begin{equation}
\bF^\ast_\varepsilon = \bF_\varepsilon\bK_\varepsilon^{-1},\;\;\;
\bX^\ast_\varepsilon = \bX + \Bkappa_\varepsilon,\;\;\;
\bK_\varepsilon = \bone +
\partial\Bkappa/\partial\bX +
\varepsilon\frac{\partial\delta\Bkappa}{\partial\bX}.
\label{variations}
\end{equation}

Standard, if lengthy, manipulations (see Appendix \ref{append})
then lead to the following set of equations governing the
stationary energy state in which the configurational variables
take on the values $\Bkappa = \Bkappa^\mathrm{s}$ and $\bK =
\bK^\mathrm{s}$.

\begin{alignat}{2}
-\mathrm{Div}^\ast\left(\psi^\ast\bone
-\bF^{\ast^\mathrm{T}}\bP^\ast +
\BSigma^\ast\right)+\frac{\partial\psi^\ast}{\partial\bX^\ast} &= \bzero &&\;\mathrm{in}\quad\Omega^\ast\label{remodequil}\\
\left( \psi^\ast\bone-\bF^{\ast^\mathrm{T}}\bP^\ast  +
\BSigma^\ast\right)\bN^\ast &= \bzero &&\;
\mathrm{on}\quad\partial\Omega^\ast_t, \label{remodbc}\\
\mbox{where}\quad\BSigma^\ast &\equiv
\frac{\partial\psi^\ast}{\partial\bK}\bK^\mathrm{T} &&\;
\mathrm{in}\quad\Omega^\ast\label{sigma}
\end{alignat}

\noindent Observe that the Eshelby stress $\psi^\ast{\bf 1} -
\bF^{\ast^\mathrm{T}}\bP^\ast$ makes its appearance. Hereafter, it
will be denoted by $\mbox{\boldmath$\sE$}$. The quantity
$\BSigma^\ast$ is a thermodynamic driving force defined in
(\ref{sigma}) as the change in Helmholtz free energy density
corresponding to a change in the tangent map of the
microstructural configuration. It is stress-like in its physical
dimensions and is a second-order tensor. For this reason we refer
to it as a non-Eshelbian configurational stress. The boundary
condition in (\ref{remodbc}) is simply a restriction on the normal
component of this thermodynamic driving term.

\subsubsection{Stationarity with simultaneous variation of
$\bu^\ast$ and $\Bkappa$}\label{sect2.1.3}

The procedure followed above is formal in the sense that the
independent imposition of variations has been assumed:
$\bu^\ast_\varepsilon = \bu^\ast +\varepsilon\delta\bu^\ast$ for
$\delta\Bkappa = \bzero$, and $\Bkappa_\varepsilon = \Bkappa +
\varepsilon\delta\Bkappa$ for $\delta\bu^\ast = \bzero$.
Physically, this may not be possible due to the interaction of
deformation and configurational changes. Stationarity of the Gibbs
free energy under simultaneous variation of its arguments is
obtained by requiring
\begin{displaymath}
\frac{\mathrm{d}}{\mathrm{d}\varepsilon}\Pi_I[\bu^\ast_\varepsilon,\Bkappa_\varepsilon]\big\vert_{\varepsilon
= 0} = 0.
\end{displaymath}

On carrying out this calculation as in Sections \ref{sect2.1.1},
\ref{sect2.1.2} and Appendix \ref{append} it can be shown that
stationarity requires the following condition:

\begin{eqnarray}
&&\int\limits_{\Omega^\ast}\left(\left(\mathrm{Div}^\ast\bP^\ast +
\bff^\ast\right)\bdot\delta\bu^\ast +
\left(\mathrm{Div}^\ast\left(\mbox{\boldmath{$\sE$}}+\BSigma^\ast\right)
+
\frac{\partial\psi^\ast}{\partial\bX^\ast}\right)\bdot\delta\Bkappa\right)\mathrm{d}V^\ast\nonumber\\
&&\hspace{1.5cm}-\int\limits_{\partial\Omega^\ast_t}\left(\left(\bP^\ast\bN^\ast
- \bar{\bt}^\ast\right)\bdot\delta\bu^\ast +
\left(\left(\mbox{\boldmath{$\sE$}}
+\BSigma\right)\bN^\ast\right)\bdot\delta\Bkappa\right)\mathrm{d}A^\ast
= 0,\label{fullstat}
\end{eqnarray}

\noindent where the notation introduced above for the Eshelby
stress and the non-Eshelbian configurational stress has been used.
Clearly (\ref{quasistatic}--\ref{quasistatic1}) and
(\ref{remodequil}--\ref{sigma}) ensure satisfaction of
(\ref{fullstat}), and are therefore sufficient conditions for
stationarity.

\section{An example of Case I remodelling: Configurational change of a long chain
molecule}\label{sect3}

In this section we employ an established statistical mechanical
example of configurational changes of long chain molecules in
order to illuminate the mathematical formulation of Section
\ref{sect2}. In addition to the interest in this example from the
standpoint of this paper we point out that configurational changes
are of central importance to the chemical activity of long chain
molecules.

\begin{figure}[ht]
\psfrag{T}{\tiny$T$} \psfrag{O}{\tiny$\frac{r_0}{2}(1-\xi)$}
\psfrag{X}{\tiny$r_0\xi$} \psfrag{K}{\tiny$\kappa$}
\psfrag{U}{\tiny$u^\ast$} \psfrag{P}{} \psfrag{Q}{}
\psfrag{L}{\tiny$\kappa+r_0$}
\includegraphics[width=12cm]{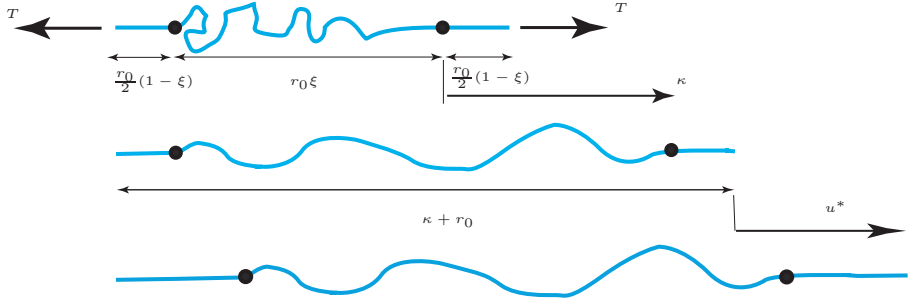}
\caption{A long chain molecule that can undergo configurational
changes and bond stretching due to an external axial load. For
ease of visualization the figure depicts bond stretching only in
the uncoiled, straight regions. } \label{figchain}
\end{figure}

Consider a long-chain molecule, say a protein, that can exist in a
highly coiled state. Let its contour length be $L$. In the
reference state, $\Omega_0$, the end-to-end lengths of the coiled
and the relatively straight domains are in the ratio $\xi\;\colon
\; 1-\xi$, and the end-to-end length of the molecule is $r_0$
(Figure \ref{figchain}). An entropic elasticity is associated with
uncoiling of the molecule. Let $\kappa$ be the change in
end-to-end length from the reference state due to uncoiling; it is
a measure of changes in the number of configurations available
(entropy). The corresponding contribution to the Gibbs free energy
function is specified by the worm-like chain (WLC) model
\citep{KratkyPorod:49}. A stiffness, $\mu$, with physical
dimensions of energy is associated with bond stretching. The
increase in length due to bond stretching is $u^\ast$. The square
of the ratio of this quantity and the length of the molecule,
$\kappa + r_0$, determines the energy stored in bonds. The
enthalpic elasticity is due to this mechanism. The molecule is
subjected to an externally-applied axial tensile force, $T$.

The one-dimensional tangent map of the local configurational
change, averaged over the entire molecule, is $K = 1 +
\kappa/r_0$. It carries the molecule to its remodelled
configuration, $\Omega^\ast$. The one-dimensional deformation
gradient relative to $\Omega^\ast$ is $F^\ast = 1 + u^\ast/(\kappa
+ r_0)$, also averaged over the molecule. It carries the molecule
to its current configuration, $\Omega$, in which it has undergone
a configurational change and bond stretching.

We will examine the response of the molecule to the axial force.
In this case, the Gibbs free energy is
\begin{eqnarray}
\Pi_I[u^\ast,\kappa] &=&  \frac{1}{2}\mu\left(\frac{u^\ast}{\kappa+r_0}\right)^2  \nonumber\\
& &+
\frac{k_B\theta}{A}\left(\frac{(\kappa+r_0)^2}{2L}+\frac{L}{4(1-(\kappa+r_0)/L)}
- \frac{\kappa+r_0}{4}\right)\nonumber\\
& &- T(u^\ast + \kappa), \label{moleculepi}
\end{eqnarray}

\noindent where the first term on the right hand-side is the
Helmholtz free energy from elastic stretching. The second term is
the entropic contribution from the WLC model, after
\citet{MarkoSiggia:95}, with $k_B$ being the Boltzmann constant
and $\theta$ being the temperature. The persistence length is $A$
and is defined as the ratio of bending stiffness to the thermal
energy. It is also the distance along the molecule's contour over
which the correlation between tangent vectors falls to $e^{-1}$.
See \citet{LandLif} for the statistical mechanics behind these
aspects of the model. The third term in (\ref{moleculepi}) is the
potential of the external force.

Proceeding as in Section \ref{sect2} we first seek stationarity
with respect to variations in $u^\ast$:
\begin{equation}
\frac{\mathrm{d}}{\mathrm{d}\varepsilon}\Pi_I[u^\ast_\varepsilon,\kappa]\Big\vert_{\varepsilon=0}
= \mu\frac{u^\ast}{(\kappa+r_0)^2} - T = 0. \label{varpiu}
\end{equation}

\noindent For this example, in the absence of a body force
(\ref{quasistatic}) reduces to the trivial requirement that the
axial force be constant along the molecule. Equation
(\ref{varpiu}) is the traction boundary condition
(\ref{quasistatic1})$_1$ for the present case. Solving, we get
\begin{equation}
{u^\ast}^\mathrm{s} = \frac{T}{\mu}(\kappa+r_0)^2, \label{solnu}
\end{equation}

\noindent where superscript $(\bullet)^s$ denotes a quantity with
respect to which the system is in a stationary state.

Next, considering variations with respect to $\kappa$ and imposing
stationarity gives
\begin{eqnarray}
\frac{\mathrm{d}}{\mathrm{d}\varepsilon}\Pi_I[u^\ast,\kappa_\varepsilon]\Big\vert_{\varepsilon=0}
&=& -\mu\frac{{u^\ast}^2}{(\kappa+r_0)^3}  - T \nonumber\\
& &+ \frac{k_B\theta}{A}\left(\frac{\kappa + r_0}{L} +
\frac{1}{4\left(1-(\kappa+r_0)/L\right)^2} -
\frac{1}{4}\right)\nonumber\\
&=& 0. \label{varpikappa}
\end{eqnarray}

In this case the differential equation (\ref{remodequil}) is
trivially satisfied since the forces corresponding to the stresses
$\mbox{\boldmath$\sE$}$ and $\BSigma^\ast$ are constant along the
molecule.

The first term in the second member of (\ref{varpikappa}) arises
due to the variation of $F^\ast$ with $\kappa$. It represents the
effect of variation in the underlying manifold, $\Omega^\ast$, due
to configurational changes. On comparison with
(\ref{intermediate}) and (\ref{result1}), and following the
derivation in Appendix \ref{append} it is clear that this term is
the reduced version of the Eshelby stress for the present
one-dimensional setting. The third term of the second member of
(\ref{varpikappa}) arises due to the variation of the WLC term
with $\kappa$. It represents the non-Eshelbian stress reduced to
this setting as is confirmed by comparison with the development in
Appendix A.

The reduced version of the boundary condition (\ref{remodbc}) is
obtained on substituting the stationary solution (\ref{solnu}) for
$u^\ast$ in  Equation (\ref{varpikappa}). (In effect, this is the
substitution referred to at the end of Appendix \ref{append}.):

\begin{equation}
-\frac{T^2}{\mu}(\kappa + r_0) +
\frac{k_B\theta}{A}\left(\frac{\kappa + r_0}{L} +
\frac{1}{4\left(1-(\kappa+r_0)/L\right)^2} - \frac{1}{4}\right) -
T = 0. \label{equilukappa}
\end{equation}

\begin{figure} \centering \psfrag{X}{\small$\kappa$ (nm)}
\psfrag{Y}{\small$D\Pi_I$ (N)}
\includegraphics[angle=0,width=8cm]{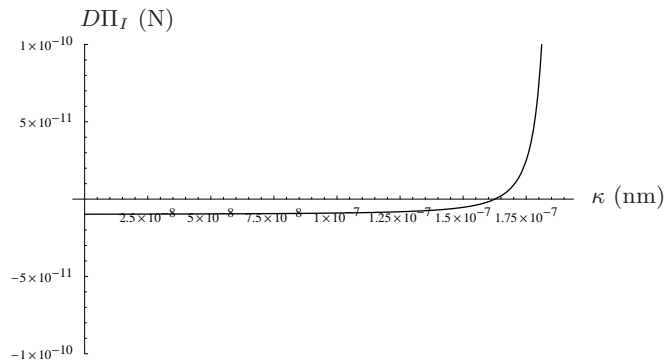}
\caption{Variation of $D\Pi_I =
(\mathrm{d}\Pi_I/\mathrm{d}\varepsilon)_{\varepsilon = 0}$ with
respect to $\kappa$.} \label{energy}
\end{figure}

Solutions to (\ref{equilukappa}), denoted by $\kappa^\mathrm{s}$,
can be obtained in closed-form since it is a cubic equation. In
order to illustrate the nature of the solution we have plotted the
left hand-side of (\ref{equilukappa}) against $\kappa$ in Figure
\ref{energy} with the numerical values of parameters in Table
\ref{table1}. The plot is restricted to the interval $0\le \kappa
< L - r_0$, since uncoiling is of interest, and (\ref{moleculepi})
is non-physical for $\kappa \ge L - r_0$.

\begin{table}[ht]
\centering \caption{Parameters used in the Gibbs free energy of
the long chain molecule}\label{table1}
\begin{tabular}{llll}
\hline
\multicolumn{1}{c}Parameter & Value & Units & Notes\\
\hline
$\xi$ & $1$ & - & - \\
$\theta$ & $300$ & K & - \\
$A$ & $14.5$ & nm & Collagen monomer molecule \citep{Sunetal:2002}\\
$r_0$ & $120$ & nm & - \\
$L$ & $309$ & nm & Collagen monomer molecule \citep{Sunetal:2002}\\
$\mu$ & $2.951\times 10^{-7}$ & J & Stiff bonds \\
$T$ & $10$ & pN & Force applied in \citet{Sunetal:2002} \\
\hline
\end{tabular}
\end{table}

The variation in free energy $D\Pi_I =
(\mathrm{d}\Pi_I/\mathrm{d}\varepsilon)_{\varepsilon = 0}$ is
negative over most of the $\kappa$-interval and passes through
zero at $\kappa^\mathrm{s} = 162.642$ nm. This is the length
increase due to uncoiling when the molecule is in the stationary
energy state. Corresponding to this value is the bond stretch,
${u^\ast}^\mathrm{s} = 3\times 10^{-9}$ nm. The molecule is very
stiff---almost rigid---to bond stretching. Also observe that the
stress-stretch response determined by the WLC model locks as
$\kappa^\mathrm{s} \to L - r_0$ from below ($\kappa^\mathrm{s} \to
189$ nm); i.e., as the uncoiled length approaches the contour
length.

\noindent\textbf{Remark 2}: The uncoiling, represented by
$\kappa$, takes place under an axial force and its effect on the
free energy is treated via a model of entropic elasticity. It
therefore follows that the entropy of the molecule is reversibly
changed by application and removal of a force.

\section{Case II remodelling: Microstructural changes represented
by internal variables}\label{sect4}

While microstructural changes always imply an evolution of the
reference configuration, a simpler description may be possible
than the general one developed in Section \ref{sect2}. Such an
instance is illustrated by the example of collagen fibril
reorientation under mechanical load: The reorientation of a fibril
does modify the reference configuration since the underlying
microstructure changes, and hence the general formulation of
Section \ref{sect2} remains applicable. However, the physics of
the process admits a simpler model that requires only the
introduction of a rotation tensor as an internal variable and is
discussed below. We begin with the biological basis and include a
description of our experiments.

\subsection{Collagen fibril alignment in gels by cell traction}
\label{sect4.1}

Collagen is the most widely-present protein in the human body. It
exists in many forms of which, to fix ideas, we will consider type
I collagen. It has a fibrillar structure and is the main form of
collagen in tendons. The fundamental unit is the triple helical
collagen molecule that is assembled from three single chain
molecular strands. The triple helix has been reported to be
between 300 to 360 nm in length and 1.5 nm in diameter. The triple
helices further assemble into collagen fibrils of varying lengths
(the range of 20--140 $\mu$m has been reported) and 10--300 nm in
diameter. The fibrils assemble into fibers that can range up to
millimeters in length. The collagen fibrils are surrounded by a
dense network of proteoglycan (PG) molecules, which, at one end,
associate with the collagen fibrils. They form a hydrated gel that
contains most of the fluid phase of the extracellular matrix.
Other extracellular matrix proteins are also found in the gel. The
interested reader is directed to \citet[chap. 19]{Alberts:2002}
and references therein for details.

\citet{StopakHarris:82} have described the reorientation of
collagen fibrils due to cell traction in gels. In their
experiments, cell-bearing tissue explants were embedded in
collagen gels. Cells were found to migrate outward from the
explant, and, by applying traction, to induce a preferred
alignment of the collagen network. Specifically, collagen fibrils
were found to align either between a pair of explants, or an
explant and a fixed, effectively rigid, support. In a
three-dimensional environment, such as a collagen gel, the
fibroblasts themselves align along directions of maximum principal
tensile stress in the matrix \citep{Balabanetal:2001}. \emph{In
vitro} (and probably \emph{in vivo} also), the stress is often
imposed by fibroblast traction on the matrix.\footnote{The aligned
fibrils stiffen the matrix in their direction, thereby providing
greater resistance to fibroblast traction. A ``positive-feedback
loop'' is thus created between fibroblast and collagen fibril
alignment, a phenomenon that is sometimes called ``contact
guidance'' \citep{BarocasTranquillo:97}.} In other \emph{in vitro}
studies an externally-applied traction has resulted in fibroblast
alignment with the maximum principal tensile stress direction.

Fibroblasts attain alignment by attaching themselves to the matrix
(usually the collagen fibrils) by ``three-dimensional adhesion
points''. Complex signalling pathways and chemical cascades are
involved in the formation of these adhesion points
\citep{Cukiermanetal:2001}. The actual attachment to the matrix is
mediated by receptors called integrins that pass through the cell
membrane and are attached to the intracellular actin network at
the opposite end from the adhesion point. See
\citet{Geigeretal:2001,Mitraetal:2005}. A tensile stress develops
in the actin network, the cell's interinsic contractile apparatus,
associated with stretching of the fibroblast between multiple
adhesion points in the extra-cellular matrix. The stress in the
actin network thus enables the fibroblast to apply traction to the
matrix in a direction now determined by the alignment of adhesion
points \citep{Balabanetal:2001}. This traction induces a marked
reorientation among the fibrils in a network with an initially
random orientation distribution. The fibrils align with the
maximum principal tensile stress direction in the matrix. There
now exists general agreement that these mutual interactions
between stress in the matrix, fibroblast alignment and stress in
the actin network are responsible for the collagen fibril
reorientation described in papers such as \citet{StopakHarris:82}.

\subsection{Experimental study}\label{exptsect}

In order to assess the role of fibroblasts in the alignment of
collagen fibrils we carried out a set of simple collagen gel
contraction experiments (that we have already alluded to in
Section \ref{sect1}).

\begin{figure}[ht]
\centering{\includegraphics[width=12cm]{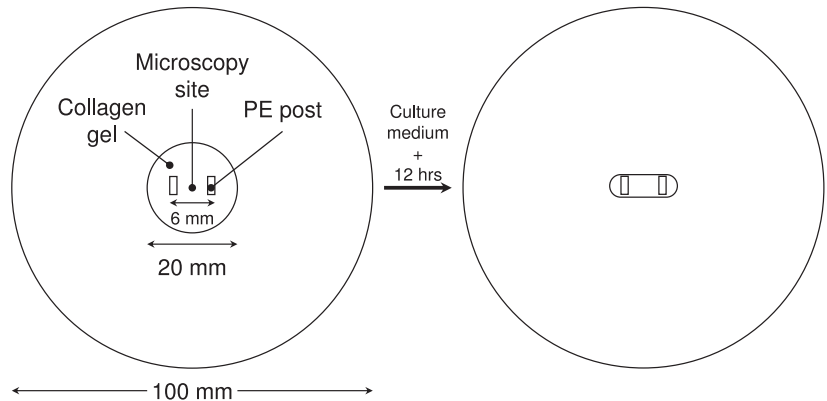}}
\caption{Schematic figure of the experimental setup for collagen
gel formation and fibril alignment by cell
traction.}\label{figplate}
\end{figure}

Three experimental configurations were considered. In the first, a
cell-collagen solution was delivered over porous polyethylene
posts spaced 6 mm apart (Figure \ref{figplate}). Over 12 hours,
the cell-seeded gel contracted to form a linear structure
stretched between the supporting posts. The contraction of the
cell-collagen gel was observed (at ambient temperature
22$^{\circ}$C) between crossed polarizing lenses mounted on an
inverted microscope for 12--15 hours under 40$\times$
magnification using an attached digital camera. An observed
increase in transmitted light through the polarizing filters over
the recording period indicated a significant microstructural
reorganization in addition to macroscopic changes in gel shape.
The gel became birefringent due to alignment of collagen fibrils
along the axis of the posts. In a second experiment, a nearly
identical procedure was followed but without the addition of the
constraining posts to the dish. While this gel contracted to half
its original diameter overnight, the lack of constraints resulted
in an isotropic contraction, no observable alignment of fibrils,
and therefore no birefringence (data not shown). Finally, a third
gel plated without cells was found to neither contract nor yield a
birefringent species. Taking these results together, one can
(reasonably) conclude that the alignment requires both
traction-providing cells (i.e., fibroblasts) as well as fixed
displacement constraints so that the stress field applied by the
cells to the matrix is anisotropic and induces the reorientation
of fibrils.

\subsection{Modelling assumptions leading to Case
II}\label{case2sect}

We make the following modelling assumptions on the alignment of
fibrils by cell traction in a collagen gel:
\begin{enumerate}
\item We restrict ourselves to situations in which there is a
uniform distribution of fibroblasts and fibrils in the gel. The
typical linear dimension of a fibroblast is 25 $\mu$m. Therefore,
at a macroscopic scale, each point in the gel has available the
microstructural mechanisms required for collagen fibril
reorientation.

\item Since the fibroblasts and fibrils are uniformly distributed, cell traction
does not result in translation of the center of mass of the
collagen fibril network within an infinitesimal neighborhood of a
point. Therefore, the only effect of fibroblast traction on the
fibrils is to change their alignment at a continuum point.

\item A consequence of Assumption 2 is
that the fluid surrounding the collagen fibrils in the gel does
not undergo a translation of its center of mass due to fibroblast
traction.

\item There is no loss of contact between the fibrils and the
surrounding fluid in the gel due to fibril realignment; i.e.,
local compatibility is maintained.
\end{enumerate}

Reverting to Equation (\ref{functnl}) we let the motion of
material points induced by fibril rotation be denoted by
$\Bkappa$. It follows from Assumptions 2 and 3 that the work done
by forces $\bff^\ast$ and $\bar{\bt}^\ast$ on $\Bkappa$ vanishes
in (\ref{functnl}) .

At the macroscopic scale, we develop an internal variable
description of the reorientation of the microscopic fibrils. In
contrast to Case I remodelling (Section \ref{sect2}) we do not
undertake a detailed consideration of the changes in reference
configuration brought about by variation of this internal
variable. The Gibbs free energy will depend on the internal
variable for fibril reorientation since it determines pointwise
anisotropy of the extracellular matrix. If the tangent to a fibril
has an initial orientation given by the unit vector field
$\bM_0(\bX) \in \Omega_0$, then $\bM^\ast(\bX,t) =
\bQ(\bX,t)\bM_0(\bX)$ is its remodelled orientation at time $t$,
where $\bQ(\bX,t) \in SO(3)$, a rotation tensor, is the internal
variable. We write the Helmholtz free energy density with respect
to $\Omega_0$ as $\hat{\psi}(\bF,\bQ,\bX)$. Note that $\hat{\psi}$
is defined with respect to the reference configuration, changes in
which we will ignore in Case II remodelling.

The Gibbs free energy functional is parametrized  by the
displacement $\bu$, now defined on $\Omega_0$, and $\bQ$.
Reflecting the assumptions made above and the arguments following
from them, Equation (\ref{functnl}) reduces to

\begin{equation}
\Pi_{II}[\bu,\bQ] =
\int\limits_{\Omega_0}\hat{\psi}(\bF,\bQ,\bX)\mathrm{d}V
-\int\limits_{\Omega_0}\bff\bdot\bu\mathrm{d}V
-\int\limits_{\partial\Omega_{0t}}\bar{\bt}\bdot\bu\mathrm{d}A.
\label{functnl1}
\end{equation}

Note that the integrals are defined on $\Omega_0$. The body force
and traction vectors are $\bff$ and $\bar{\bt}$, respectively,
defined per unit volume in $\Omega_0$ and per unit surface area on
$\partial\Omega_0$. We now consider variations on $\bQ$, recalling
that $SO(3)$ is a Lie group whose Lie algebra consists of real,
skew-symmetric matrices. This Lie algebra is denoted by
$\mathrm{Skw}(3)$ for the case wherein $\bQ$ acts on vectors in
$\mathbb{R}^3$ (see any standard textbook on manifold analysis,
for instance, \citet{Choquet-Bruhat}). The admissible variations
on $\bQ$ are therefore of the form
\begin{equation}
\bQ_\varepsilon = \bQ + \delta\bQ,\;\;\mathrm{where}\;\delta\bQ =
\bW\bQ\,\forall\,\bW\in\mathrm{Skw}(3). \label{liealgebra}
\end{equation}

\noindent This leads to the following variation on $\Pi_{II}$:
\begin{eqnarray}
&&\frac{\mathrm{d}}{\mathrm{d}\varepsilon}\Pi_{II}[\bu,\bQ_\varepsilon]\big\vert_{\varepsilon=0}=\nonumber\\
&&\qquad\qquad\frac{\mathrm{d}}{\mathrm{d}\varepsilon}\left(\int\limits_{\Omega}\hat{\psi}(\bF,\bQ_\varepsilon,\bX)\mathrm{d}V
-\int\limits_{\Omega}\bff\bdot\bu\mathrm{d}V -
\int\limits_{\partial\Omega_t}\bar{\bt}\bdot\bu\mathrm{d}A\right)\Big\vert_{\varepsilon=0}.
\end{eqnarray}

\noindent Imposing stationarity, invoking the localization
argument and applying (\ref{liealgebra}) this reduces in a
straightforward manner to
\begin{equation}
\frac{\partial\psi}{\partial\bQ}\bcolon\bW\bQ =
0,\;\;\forall\;\bW\in\mathrm{Skw}(3),
\end{equation}

\noindent a relation also obtained by \citet{Vianello:96}, albeit
without beginning from the integral form. As a final step this
result for a stationary state can be rewritten as
\begin{equation}
\frac{\partial\psi}{\partial\bQ}\bQ^\mathrm{T}\Big\vert_{\bQ=\bQ^\mathrm{s}}\bcolon\bW
= 0,\;\;\forall\;\bW\in\mathrm{Skw}(3). \label{qequil}
\end{equation}

\subsubsection{Example: Collagen fibril reorientation in a gel using a
continuum strain energy function} \label{sect4.1.1}

The WLC model (\ref{moleculepi}) is extended to a continuum strain
energy function for the system consisting of the gel, collagen
fibrils and fibroblasts. This is achieved by introduction of the
fibril number density, $N$, and addition of a repulsive term to
enforce a vanishing stress at unit stretch and a bulk
compressibility term for the gel:

\begin{eqnarray}
\hat{\psi}(\bF,\bQ,\bX) &=& \frac{N k_B \theta}{4
A}\left(\frac{2r^2}{L} + \frac{L}{1-r/L} -
r\right)\nonumber\\
&&\underbrace{-\frac{N k_B \theta}{4A}\left(\frac{1}{L} +
\frac{1}{4r_0(1 - \frac{r_0}{L})^2} -\frac{1}{4r_0}
\right)\log(\lambda^{4r_0^2})}_{\mbox{repulsive term}}\nonumber\\
&&\underbrace{+ \frac{\gamma}{\beta}(J^{-2\beta} -1) + 2\gamma{\bf
1}\bcolon\bE}_{\mbox{bulk compressibility}}. \label{wlcmeq}
\end{eqnarray}

The elastic stretch in the direction of the remodelled fibril is
$\lambda$, and $\bE = \frac{1}{2}(\bF^\mathrm{T}\bF - \bone)$ is
the Lagrange strain. The factors $\gamma$ and $\beta$ control bulk
compressibility. The end-to-end length is given by

\begin{equation}
r = \sqrt{r_0^2\lambda^2},\quad \lambda =
\sqrt{\bM^\ast\bdot\bF^\mathrm{T}\bF\bM^\ast}. \label{rwlcm}
\end{equation}

\noindent Note that the end-to-end length is modified by the
macroscopic deformation through (\ref{rwlcm}). This is in contrast
with the explicit inclusion of the change in molecular
configuration, $\kappa$, in (\ref{moleculepi}). The change in
configuration considered in the present section is at the more
macroscopic scale of fibril reorientation.

Consider a cylindrically-shaped collagen gel in which the collagen
fibrils are all radially-oriented at time $t=0$. Adopting a
cylindrical coordinate system, $\{\be_R,\be_\alpha,\be_Z\}$, we
have $\bM_0 = \be_R$. The gel is loaded by imposing a deformation
gradient: $\bF = \lambda_R\be_R\otimes\be_R +
\lambda_\alpha\be_\alpha\otimes\be_\alpha +
\lambda_Z\be_Z\otimes\be_Z$, such that $\lambda_Z > 1$, $\lambda_R
= \lambda_\alpha$ and $\lambda_Z > \lambda_R$. We will assume that
Case II remodelling occurs according to the cell traction
mechanisms discussed above, and that the fibrils undergo local
reorientation with $\bM^\ast \to \be_Z$ as $t \to \infty$. The
rotation tensor corresponding to this stationary state is
$\bQ^\mathrm{s} = -\be_R\otimes\be_Z + \be_\alpha\otimes\be_\alpha
+ \be_Z\otimes\be_R$. We aim to verify that $\bQ^\mathrm{s}$
satisfies (\ref{qequil}).

From (\ref{wlcmeq}) and (\ref{rwlcm}) we have, after some
manipulations,
\begin{eqnarray}
&
&\frac{\partial\psi}{\partial\bQ}\bQ^\mathrm{T}\Big\vert_{\bQ=\bQ^\mathrm{s}}\bcolon\bW
=\nonumber\\
& &\quad\frac{N k_B\theta
r_0}{4A}\left[\left(\frac{4r}{L}+\frac{1}{(1-r/L)^2}-1\right)
-\left(\frac{4 r_0^2}{rL} + \frac{r_0}{r(1-r_0/L)^2} -
\frac{r_0}{r}\right)\right]\nonumber\\
& &
\hspace{5cm}\cdot\bF^\mathrm{T}\bF(\bQ^\mathrm{s}\bM_0\otimes\bQ^\mathrm{s}\bM_0)\bcolon\bW
\label{qequil1}
\end{eqnarray}

\noindent Using the tensor product expansions established above
for $\bF$ and $\bQ^\mathrm{s}$, and $\bM_0 = \be_R$, this reduces
to
\begin{eqnarray}
&
&\frac{\partial\psi}{\partial\bQ}\bQ^\mathrm{T}\Big\vert_{\bQ=\bQ^\mathrm{s}}\bcolon\bW
=\nonumber\\
& &\quad\frac{N k_B\theta
r_0}{4A}\left[\left(\frac{4r}{L}+\frac{1}{(1-r/L)^2}-1\right)-
\left(\frac{4 r_0^2}{rL} + \frac{r_0}{r(1-r_0/L)^2} -
\frac{r_0}{r}\right)\right]\nonumber\\
& &\hspace{7.0cm}\lambda_Z^2\be_Z\otimes\be_Z\bcolon\bW.
\label{qequil2}
\end{eqnarray}

Since (\ref{qequil2}) involves the scalar product of a symmetric
tensor and $\bW\in\mathrm{Skw}(3)$, we have
\begin{eqnarray}
\frac{\partial\psi}{\partial\bQ}\bQ^\mathrm{T}\Big\vert_{\bQ=\bQ^\mathrm
{s}}\bcolon\bW
=0,\quad&&\forall\bW\in\mathrm{Skw}(3),\nonumber\\
&&\bQ^\mathrm{s} = -\be_R\otimes\be_Z +
\be_\alpha\otimes\be_\alpha + \be_Z\otimes\be_R. \label{qequil3}
\end{eqnarray}

This implies that a stationary energy state is achieved when the
fibrils with initial radial orientation undergo reorientation
along the direction of maximum principal tensile stretch.

\noindent\textbf{Remark 3}: Even though the discussion in Sections
\ref{sect4.1} and \ref{exptsect} refers to fibril alignment with
the maximum principal tensile stress direction, it is entirely
equivalent to consider alignment with the maximum principal
stretch direction in the stationary energy state. This is on
account of the work of \citet{Vianello:96} who showed that the
strain energy of an anisotropic solid is at a minimum when the
principal stress and stretch directions coincide. This result is
used to justify an evolution law for fibril orientation
(\ref{rateeq1}) below.

\subsection{Notes on a recent model of fiber orientation}\label{sect4.2}

To conclude this section, we briefly discuss a recent theory of
remodelling, developed by \citet{Driessenetal1:03}, for the
evolution of fiber orientation within a tissue. Their theory is
quite distinct from the development in Sections
\ref{sect4.1}--\ref{case2sect}, and has at least two fundamental
shortcomings:

\begin{itemize}
\item[(\romannumeral 1)] In order to represent the distribution of
fiber orientation, these authors work with a fiber orientation
tensor, defined as $\bS_0 = \langle \be_0,\be_0\rangle :=
\sum_{i=1}^N \psi_i \be_0^i\otimes\be_0^i$. There are $N$ distinct
orientations, each specified by a unit vector, $\be_0^i$, and with
a probability distribution $\psi_i$, such that $\sum_{i=1}^N
\psi_i = 1$. The tensor, $\bS_0$, however fails to distinguish
between cubic orthotropy ($N = 3$, $\psi_i = 1/3$,
$\be_0^i\bdot\be_0^j = \delta^{ij}$ the Kronecker-delta) and
isotropy ($\psi_i = 1/N$, the directions $\be_0^i$ being
distributed uniformly over the unit sphere, in the limit $N \to
\infty$). A straightforward calculation shows that $\bS_0 =
\frac{1}{3}\bone$ in both cases. This fundamental failing suggests
that higher-order statistics must be included, for instance,
moments of the fiber distribution.

\item[(\romannumeral 2)] The
second limitation is related to the evolution law prescribed for
the fiber orientation tensor in the current placement, $\bS =
(1/\Lambda^2)\bF\bS_0\bF^\mathrm{T}$, where $\Lambda^2$ is the
mean square of the stretches of the fibers:
\begin{displaymath}
\Lambda^2 =
\sum_{i=1}^N\psi^i\be_0^i\bdot\bF^\mathrm{T}\bF\be_0^i.
\end{displaymath}

The evolution law prescribed is $\stackrel{\nabla}{\bS} +
2(\bD\bcolon\bS)\bS = \frac{1}{\tau}(\bA - \bS)$, where
$\stackrel{\nabla}{\bS}$ is the Lie derivative of $\bS$, $\bD$ is
the rate of deformation tensor, $\tau$ is the relaxation time, and
$\bA$ is written as a function of the finger tensor: $\bA =
\bB^\nu/\mathrm{tr}(\bB^\nu)$, with $\nu$ being a real exponent.
The physical implication of such a law is that $\bS$ continues to
evolve, driven by the strain, until $\bS =
\bB^\nu/\mathrm{tr}(\bB^\nu)$: For a point with $\bB = \bone$,
i.e., unchanged from its reference placement, this means that
\emph{any} initial fiber orientation tensor, $\bS_0(0) = \bS(0)$
(e.g., $\bS_0(0) = \bS(0) = \be_0\otimes\be_0$) must evolve until
$\bS_0(t) = \bS(t) = \bone$. We know of no physiological instances
in which this happens. In fact, this conclusion does not
correspond with common experience. In our experiments described in
Section \ref{exptsect} we have also confirmed that the fiber
orientations in a uniaxially-aligned collagen gel remain in this
state if the gel does not deform, and do not evolve to $\bS(t) =
\bone$ as Driessen and co-workers' rule suggests.
\end{itemize}

\section{Thermodynamic dissipation associated with reorienting
fibrils} \label{sect4.1.2}

We return to our formulation of collagen fibril re-orientation via
internal variables for the discussion on dissipation. Denoting the
internal energy of the system (gell, collagen fibrils and
fibroblasts) by $e$, and the energy lost against viscous
resistance by $\mathscr{D}_v$, the First Law of thermodynamics
gives
\begin{equation}
\dot{e} = \bP\bcolon\dot{\bF} - \mathscr{D}_v - \Bnabla\bdot\bq,
\label{firstlaw}
\end{equation}

\noindent where $\bP$ is the first Piola-Kirchhoff stress defined
on $\Omega_0$, and $\bq$ is the heat flux. The energy lost against
viscous resistance as the fibrils rotate relative to the viscous
gel is in the form of heat, therefore $-\mathscr{D}_v$ is heat
lost from the system. To be more precise we assume a linear
viscosity and write $\mathscr{D}_v =
\frac{1}{2}\mu\vert\Bomega\vert^2$, where $\mu$ is the viscosity
of the gel, and $\Bomega =
-\frac{1}{2}\Bepsilon\bcolon(\dot{\bQ}\bQ^\mathrm{T})$ is the
angular velocity of the fibrils with $\Bepsilon$ being the
permutation symbol.

At steady temperature (an isothermal process, such as that in our
experiments of Section \ref{exptsect}) we have, from a Legendre
transformation, $\dot{e} = \dot{\psi} + \theta\dot{\eta}$, where
$\eta$ is the entropy density of the system. In anticipation of
the arguments to follow we now write the Helmholtz free energy
density as a sum of mechanical and chemical components, $\psi =
\psi_m + \psi_c$ where $\psi_m = \hat{\psi}_m(\bF,\bQ,\bX)$ is
given by (\ref{wlcmeq}), which was written for a purely mechanical
system. Therefore, the internal energy density rate satisfies
\begin{equation}
\dot{e} = \dot{\psi}_m + \dot{\psi}_c + \theta\dot{\eta}.
\label{legendre}
\end{equation}

The entropy density rate is governed by the Second Law of
thermodynamics, which we write first as an equality:
\begin{equation}
\dot{\eta} = -\frac{\mathscr{D}_v}{\theta}
-\Bnabla\bdot\left(\frac{\bq}{\theta}\right) + \gamma,
\label{secondlaw}
\end{equation}

\noindent where the first term on the right hand-side is the
entropy loss due to the heat sink, the second term is the entropy
change due to the heat flux, and $\gamma$ is an entropy production
term due to irreversible processes internal to the system. The
statement of the Second Law as an inequality is
\begin{equation}
\gamma \ge 0. \label{secondlawineq}
\end{equation}

Multiplying (\ref{secondlaw}) by $\theta$, and combining it with
(\ref{firstlaw}) and (\ref{legendre}), we have
\begin{displaymath}
\dot{\psi}_m + \dot{\psi}_c -\bP\bcolon\dot{\bF} +
\frac{\bq\bdot\Bnabla\theta}{\theta} + \theta\gamma = 0.
\end{displaymath}

\noindent Using $\psi_m = \hat{\psi}_m(\bF,\bQ,\bX)$ as justified
above, this can be expanded to
\begin{displaymath}
\frac{\partial\psi_m}{\partial\bF}\bcolon\dot{\bF} +
\frac{\partial\psi_m}{\partial\bQ}\bcolon\dot{\bQ} + \dot{\psi}_c
-\bP\bcolon\dot{\bF} + \frac{\bq\bdot\Bnabla\theta}{\theta} +
\theta\gamma = 0.
\end{displaymath}

\noindent The general hyperelastic constitutive law $\bP =
\partial\psi_m/\partial\bF$ reduces this inequality to
\begin{equation}
\frac{\partial\psi_m}{\partial\bQ}\bcolon\dot{\bQ} + \dot{\psi}_c
+ \frac{\bq\bdot\Bnabla\theta}{\theta} + \theta\gamma = 0.
\label{dissip1}
\end{equation}

Using (\ref{wlcmeq}) and (\ref{rwlcm}) with
$\hat{\psi}(\bF,\bQ,\bX)$ replaced by $\hat{\psi}_m(\bF,\bQ,\bX)$
for the first term in (\ref{dissip1}), we have

\begin{eqnarray}
&&\frac{N k_B\theta
r_0}{4A}\left[\left(\frac{4r}{L}+\frac{1}{(1-r/L)^2}-1\right)
-\left(\frac{4 r_0^2}{rL} + \frac{r_0}{r(1-r_0/L)^2} -
\frac{r_0}{r}\right)\right]\nonumber\\
& &
\hspace{2.5cm}\cdot\bF^\mathrm{T}\bF(\bQ\be_R\otimes\dot{\bQ}\be_R)\bcolon\bone
+ \dot{\psi}_c + \frac{\bq\bdot\Bnabla\theta}{\theta} +
\theta\gamma = 0, \label{dissip2}
\end{eqnarray}

\noindent where $\bM^\ast = \bQ\bM_0$ and $\bM_0 = \be_R$ have
been used. In a recent paper, \citet{kuhletal_orient:04} have
proposed the following first-order rate equation for the local
evolution of fibril orientation:
\begin{equation}
\frac{\partial\bM^\ast}{\partial t} =
-\frac{1}{\tau}\left[\left(\bM^\ast\bdot\bM_\mathrm{max}\right)\bM^\ast
- \bM_\mathrm{max}\right], \label{rateeq1}
\end{equation}

\noindent where $\tau > 0$ is a relaxation time, and
$\bM_\mathrm{max}$ is the eigen vector corresponding to the
maximum principal tensile stretch (see Remark 3). For the example
in Section \ref{sect4.1.1} we have $\bM_\mathrm{max} = \be_Z$.
Equation (\ref{rateeq1}) can be written in terms of $\bQ$ by
substituting $\bM^\ast = \bQ\be_R$:
\begin{equation}
\dot{\bQ}\be_R =
-\frac{1}{\tau}\left[\left(\be_Z\bdot\bQ\be_R\right)\bQ\be_R -
\be_Z\right]. \label{rateeq2}
\end{equation}

Combining (\ref{rateeq2}) and (\ref{dissip2}) we obtain
\begin{eqnarray}
&&-\frac{N k_B\theta
r_0}{4A}\left[\left(\frac{4r}{L}+\frac{1}{(1-r/L)^2}-1\right)
-\left(\frac{4 r_0^2}{rL} + \frac{r_0}{r(1-r_0/L)^2} -
\frac{r_0}{r}\right)\right]\nonumber\\
& &
\cdot\bF^\mathrm{T}\bF\left\{\bQ\be_R\otimes\frac{1}{\tau}\left[\left(\be_Z\bdot\bQ\be_R\right)\bQ\be_R-\be_Z\right]\right\}\bcolon\bone
\nonumber\\
&&+ \dot{\psi}_c + \frac{\bq\bdot\Bnabla\theta}{\theta}+
\theta\gamma = 0. \label{dissip3}
\end{eqnarray}

\noindent Returning to the tensor product expansion for $\bF$, we
use $\bone = \be_R\otimes\be_R + \be_\alpha\otimes\be_\alpha +
\be_Z\otimes\be_Z$, and the fact that, for uniaxial tension,
$\lambda_R = \lambda_\alpha$ to write $\bF^\mathrm{T}\bF =
\lambda_R^2\bone + (\lambda_Z^2 - \lambda_R^2)\be_Z\otimes\be_Z$.
Using this relation in (\ref{dissip3}) we get, after some
manipulations,
\begin{eqnarray}
&&-\frac{N k_B\theta
r_0}{4A}\left[\left(\frac{4r}{L}+\frac{1}{(1-r/L)^2}-1\right)
-\left(\frac{4 r_0^2}{rL} + \frac{r_0}{r(1-r_0/L)^2} -
\frac{r_0}{r}\right)\right]\nonumber\\
& & \hspace{1cm}\cdot
\frac{\lambda_Z^2}{\tau}\left(\be_Z\bdot\bQ\be_R\right)\left[\left(\be_Z
\bdot\bQ\be_R\right)^2 -1\right] + \dot{\psi}_c +
\frac{\bq\bdot\Bnabla\theta}{\theta}+ \theta\gamma = 0.
\label{dissip4}
\end{eqnarray}

\noindent From (\ref{rateeq2}) we have
\begin{equation}
\be_Z\bdot\dot{\bQ}\be_R =
-\frac{1}{\tau}\left[\left(\be_Z\bdot\bQ\be_R\right)^2-
1\right]\ge 0 \label{rateeq3}
\end{equation}

\noindent At $t = 0$ the internal variable $\bQ(t) = \bone$,
leading to $\be_Z\bdot\bQ(0)\be_R = 0$ . From this and
(\ref{rateeq3}) it follows that $\be_Z\bdot\bQ(t)\be_R > 0$ for
all $t > 0$. Furthermore, for any locally remodelled orientation
$\bM^\ast = M_Z \be_Z + M_R \be_R$, the unstretched and stretched
lengths of the fibrils are in the ratio $r_0/r \le 1$ provided
$(M_Z\lambda_Z)^2 + (M_R\lambda_R)^2 > 1$. Using these results the
first term in (\ref{dissip4}) can be shown to be greater than zero
provided the fibrils have rotated sufficiently for
$(M_Z\lambda_Z)^2 + (M_R\lambda_R)^2
> 1$ to hold, and $r < L$.

This conclusion regarding the first term in (\ref{dissip4}) is not
surprising. Its interpretation is that, for a given state of
deformation $\bF$, the strain energy density at a point increases
as $\bQ$ locally rotates fibrils to align with the direction of
maximum principal tensile stretch \citep[see][]{Arrudaetal:2005}.
In fact, this parametric variation is a property of any
fiber-reinforced material. The contributions of the remaining
terms in (\ref{dissip4}) will now be studied towards understanding
the dissipative character of biological remodelling.

From the Second Law written as an inequality
(\ref{secondlawineq}), it follows that the entropy production term
is positive semi-definite. Assuming Fourier's Law of heat
conduction, $\bq = -\bK_\mathrm{con}\Bnabla\theta$, where
$\bK_\mathrm{con}$ is a positive semi-definite heat conduction
tensor, ensures that heat conduction results in a non-positive
dissipation. Finally, since, as shown by our experiments (Section
\ref{exptsect}) re-orientation of fibrils happens by cell
traction, the cells must consume their chemical free energy in
this process: $\dot{\psi}_c \le 0$. Rewriting (\ref{dissip4}) with
a reduced form for the first term on the left hand-side,
\begin{equation}
\underbrace{\frac{\partial\psi_m}{\partial\bQ}\bcolon\dot{\bQ}}_{\ge
0} + \underbrace{\dot{\psi}_c}_{\le 0} +
\underbrace{\frac{\bq\bdot\Bnabla\theta}{\theta}}_{\le 0}+
\underbrace{\theta\gamma}_{\ge 0} = 0.\label{dissip5}
\end{equation}

Written in this form, the dissipation has several implications:
\begin{enumerate}
\item Perhaps the most significant implication from the standpoint
of development of models is that a purely mechanical theory is
thermodynamically inadmissible for remodelling processes that
stiffen the material. This is seen from (\ref{dissip5}) without
the second and third terms, since the left hand-side is
necessarily greater than zero if remodelling takes place. Either
heat conduction or the chemical free energy changes must be
considered to satisfy the dissipation relation.

\item The chemical and mechanical action of cells needs to be considered in quantitative
terms. While the chemical term is represented by $\psi_c$ a
constitutive model is still needed for it. Intra-cellular
mechanical changes are also involved in the process (Section
\ref{exptsect}), and must be modelled by a separate free energy
term.

\item The changes in chemical and mechanical free energy of the
cells translates to entropy changes also. At this stage it is
unclear whether the cells lower or increase their entropy by these
processes. If the answer is a decrease, the overall positive
entropy change must be sought further afield.

\item While it is common to model biological tissues as
isothermal, and even to ignore heat transport in them,
(\ref{dissip5}) suggests that this is not necessarily allowable.
Indeed it now appears that some heat flux must exist in the
tissues for transport of the energetic and entropic byproducts of
cellular activity. This term is indispensable if chemcial free
energy is ignored.
\end{enumerate}

\section{Discussion and conclusion}\label{sect7}

The hypothesis that biological systems attain stationary energy
states with respect to changes in their microstructure is examined
in this paper. A variational treatment can be applied to result in
at least two, quite different, descriptions of remodelling: The
first involving an evolution of the underlying material
configuration, and the second described by internal variables. We
note that the range of phenomena that can be described span from
molecular to tissue scales.

While the notion of stationary states of the free energy is
examined, it is clear that dissipation must play a central role.
This is highlighted by Section \ref{sect4.1.2} and its result that
stiffening of active biological materials requires both:
consideration of chemical free energy and an entropic sink. This
is a conclusion that merits deeper study. Attainment of true
equilibrium states, in which all processes cease is dictated by
dissipation. For this reason, and because biological systems have
many mechanisms by which to dissipate energy, a very careful
consideration of the Second Law and its consequences is essential
to studies of remodelling.

\appendix
\section{Variational calculus for Case I remodelling}
\label{append}

Letting $J^\ast$ denote $\mathrm{det}(\bF^\ast)$, $\psi$ denote
the Helmholtz free energy density with respect to the reference
configuration, and $\bff$ denote the body force in the reference
configuration, we rewrite the integrals in (\ref{varint}) over the
reference configuration using $J^\ast\psi^\ast = \psi$ and
$J^\ast\bff^\ast = \bff$.
\begin{eqnarray*}
\frac{\mathrm{d}}{\mathrm{d}\varepsilon}\Pi[\bu^\ast,\Bkappa_\varepsilon]\Big\vert
&=&
\int\limits_{\Omega_0}\left(\frac{\mathrm{d}J^\ast_\varepsilon}{\mathrm{d}\varepsilon}\psi^\ast_\varepsilon
+
J^\ast_\varepsilon\frac{\mathrm{d}\psi^\ast_\varepsilon}{\mathrm{d}\varepsilon}\right)\mathrm{d}V\\
& &
-\int\limits_{\Omega_0}J^\ast\bff^\ast\cdot\frac{\mathrm{d}\Bkappa_\varepsilon}{\mathrm{d}\varepsilon}\mathrm{d}V
-
\int\limits_{\partial\Omega_{0t}}\bar{\bt}\cdot\frac{\mathrm{d}\Bkappa_\varepsilon}{\mathrm{d}\varepsilon}\mathrm{d}A.
\end{eqnarray*}

\noindent Reverting to the remodelled configuration,
\begin{eqnarray*}
\frac{\mathrm{d}}{\mathrm{d}\varepsilon}\Pi[\bu^\ast,\Bkappa_\varepsilon]
&=&
\int\limits_{\Omega^\ast}J^{\ast^{-1}}\left(\frac{\mathrm{d}J^\ast_\varepsilon}{\mathrm{d}\varepsilon}\psi^\ast_\varepsilon
+
J^\ast_\varepsilon\frac{\mathrm{d}\psi^\ast_\varepsilon}{\mathrm{d}\varepsilon}\right)\mathrm{d}V^\ast\\
& &
-\int\limits_{\Omega^\ast}\bff^\ast\cdot\frac{\mathrm{d}\Bkappa_\varepsilon}{\mathrm{d}\varepsilon}\mathrm{d}V^\ast
-\int\limits_{\partial\Omega^\ast_t}\bar{\bt}^\ast\cdot\frac{\mathrm{d}\Bkappa_\varepsilon}{\mathrm{d}\varepsilon}\mathrm{d}A^\ast.
\end{eqnarray*}

\noindent Applying the chain rule and invoking stationarity with
respect to variations in $\Bkappa$,
\begin{eqnarray}
\frac{\mathrm{d}}{\mathrm{d}\varepsilon}\Pi[\bu^\ast,\Bkappa_\varepsilon]\Big\vert_{\varepsilon=0}
&=&
\left(\int\limits_{\Omega^\ast}J^{\ast^{-1}}\left(\frac{\partial
J^\ast}{\partial
\bK}\colon\frac{\mathrm{d}\bK_\varepsilon}{\mathrm{d}\varepsilon}\psi^\ast_\varepsilon+
J^\ast_\varepsilon\frac{\partial\psi^\ast}{\partial
\bF^\ast}\colon\frac{\mathrm{d}\bF^\ast_\varepsilon}{\mathrm{d}\varepsilon}\right)\mathrm{d}V^\ast\right)_{\varepsilon=0}\nonumber\\
&
&+\left(\int\limits_{\Omega^\ast}J^{\ast^{-1}}\left(J^\ast_\varepsilon\frac{\partial\psi^\ast}{\partial
\bK}\colon\frac{\mathrm{d}\bK_\varepsilon}{\mathrm{d}\varepsilon}
+J^\ast_\varepsilon\frac{\partial\psi^\ast}{\partial\bX^\ast}\colon\frac{\mathrm{d}\Bkappa_\varepsilon}{\mathrm{d}\varepsilon}\right)\mathrm{d}V^\ast\right)_{\varepsilon=0}\nonumber\\
&
&-\left(\int\limits_{\Omega^\ast}\bff^\ast\cdot\frac{\mathrm{d}\Bkappa_\varepsilon}{\mathrm{d}\varepsilon}\mathrm{d}V^\ast+\int\limits_{\partial\Omega^\ast_t}\bar{\bt}^\ast\cdot\frac{\mathrm{d}\Bkappa_\varepsilon}{\mathrm{d}\varepsilon}\mathrm{d}A^\ast\right)_{\varepsilon=0}.
\label{intermediate}
\end{eqnarray}

The derivatives with respect to $\varepsilon$ in
(\ref{intermediate}) are obtained from (\ref{variations}):
\begin{equation}
\left(\frac{\mathrm{d}\bK_\varepsilon}{\mathrm{d}\varepsilon}\right)_{\varepsilon
= 0} =
\frac{\partial\delta\Bkappa}{\partial\bX},\;\;\;\left(\frac{\mathrm{d}\bF^\ast_\varepsilon}{\mathrm{d}\varepsilon}\right)_{\varepsilon
= 0} = (\bone -
\bF^\ast)\frac{\partial\delta\Bkappa}{\partial\bX^\ast},\;\;\;
\left(\frac{\mathrm{d}\Bkappa_\varepsilon}{\mathrm{d}\varepsilon}\right)_{\varepsilon
= 0} = \delta\Bkappa.\label{result1}
\end{equation}

\noindent Substituting (\ref{result1}) in (\ref{intermediate})
gives
\begin{eqnarray*}
\frac{\mathrm{d}}{\mathrm{d}\varepsilon}\Pi[\bu^\ast,\Bkappa_\varepsilon]\Big\vert_{\varepsilon=0}
&=& \int\limits_{\Omega^\ast}J^{\ast^{-1}}\left(\frac{\partial
J^\ast}{\partial
\bK}\colon\frac{\partial\delta\Bkappa}{\partial\bX}\psi^\ast+
J^\ast\frac{\partial\psi^\ast}{\partial \bF^\ast}\colon(\bone -
\bF^\ast)\frac{\partial\delta\Bkappa}{\partial\bX^\ast}\right)\mathrm{d}V^\ast\\
&&+\int\limits_{\Omega^\ast}J^{\ast^{-1}}\left(J^\ast\frac{\partial\psi^\ast}{\partial
\bK}\colon\frac{\partial\delta\Bkappa}{\partial\bX}
+J^\ast\frac{\partial\psi^\ast}{\partial\bX^\ast}\delta\Bkappa\right)\mathrm{d}V^\ast\\
&
&-\int\limits_{\Omega^\ast}\bff^\ast\cdot\delta\Bkappa\mathrm{d}V^\ast-\int\limits_{\partial\Omega^\ast_t}\bar{\bt}^\ast\cdot\delta\Bkappa\mathrm{d}A^\ast
\end{eqnarray*}

\noindent Using $\partial J^\ast/\partial\bK :=
\partial[\mathrm{det}(\bK)]/\partial\bK =
\mathrm{det}(\bK)\bK^{-\mathrm{T}}$, this reduces to
\begin{eqnarray*}
\frac{\mathrm{d}}{\mathrm{d}\varepsilon}\Pi[\bu^\ast,\Bkappa_\varepsilon]\Big\vert_{\varepsilon=0}
&=&
\int\limits_{\Omega^\ast}\left(\bK^{-\mathrm{T}}\colon\frac{\partial\delta\Bkappa}{\partial\bX}\psi^\ast+
\frac{\partial\psi^\ast}{\partial \bF^\ast}\colon(\bone -
\bF^\ast)\frac{\partial\delta\Bkappa}{\partial\bX^\ast}\right)\mathrm{d}V^\ast\\
&&+\int\limits_{\Omega^\ast}\left(\frac{\partial\psi^\ast}{\partial
\bK}\colon\frac{\partial\delta\Bkappa}{\partial\bX}
+\frac{\partial\psi^\ast}{\partial\bX^\ast}\delta\Bkappa\right)\mathrm{d}V^\ast\\
&
&-\int\limits_{\Omega^\ast}\bff^\ast\cdot\delta\Bkappa\mathrm{d}V^\ast-\int\limits_{\partial\Omega^\ast_t}\bar{\bt}^\ast\cdot\delta\Bkappa\mathrm{d}A^\ast
\end{eqnarray*}

\noindent Using standard manipulations of the scalar product of
tensors, and the chain rule, $\partial(\bullet)/\partial\bX^\ast =
(\partial(\bullet)/\partial\bX)\bK^{-1}$, the preceding equation
yields
\begin{eqnarray*}
\frac{\mathrm{d}}{\mathrm{d}\varepsilon}\Pi[\bu^\ast,\Bkappa_\varepsilon]\Big\vert_{\varepsilon=0}
&=&
\int\limits_{\Omega^\ast}\left(\psi^\ast\bone\colon\frac{\partial\delta\Bkappa}{\partial\bX^\ast}+
(\bone - \bF^{\ast^\mathrm{T}})\frac{\partial\psi^\ast}{\partial
\bF^\ast}\colon\frac{\partial\delta\Bkappa}{\partial\bX^\ast}\right)\mathrm{d}V^\ast\\
&&+\int\limits_{\Omega^\ast}\left(\frac{\partial\psi^\ast}{\partial
\bK}\bK^\mathrm{T}\colon\frac{\partial\delta\Bkappa}{\partial\bX^\ast}
+\frac{\partial\psi^\ast}{\partial\bX^\ast}\delta\Bkappa\right)\mathrm{d}V^\ast\\
&
&-\int\limits_{\Omega^\ast}\bff^\ast\cdot\delta\Bkappa\mathrm{d}V^\ast-\int\limits_{\partial\Omega^\ast_t}\bar{\bt}^\ast\cdot\delta\Bkappa\mathrm{d}A^\ast
\end{eqnarray*}

Defining the configurational stress, $\BSigma^\ast :=
(\partial\psi^\ast/\partial\bK)\bK^\mathrm{T}$, and introducing
the first Piola-Kirchhoff stress, this can be written as
\begin{eqnarray*}
& &
\frac{\mathrm{d}}{\mathrm{d}\varepsilon}\Pi[\bu^\ast,\Bkappa_\varepsilon]\Big\vert_{\varepsilon=0}
=\\
& &\qquad\qquad\qquad
\int\limits_{\Omega^\ast}\left(\psi^\ast\bone\colon\frac{\partial\delta\Bkappa}{\partial\bX^\ast}+
(\bone -
\bF^{\ast^\mathrm{T}})\bP^\ast\colon\frac{\partial\delta\Bkappa}{\partial\bX^\ast}
+\BSigma^\ast\colon\frac{\partial\delta\Bkappa}{\partial\bX^\ast}\right)\mathrm{d}V^\ast\\
&
&\qquad\qquad\qquad+\int\limits_{\Omega^\ast}\frac{\partial\psi^\ast}{\partial\bX^\ast}\delta\Bkappa\mathrm{d}V^\ast-\int\limits_{\Omega^\ast}\bff^\ast\cdot\delta\Bkappa\mathrm{d}V^\ast-\int\limits_{\partial\Omega^\ast_t}\bar{\bt}^\ast\cdot\delta\Bkappa\mathrm{d}A^\ast
\end{eqnarray*}

\noindent The Divergence Theorem allows this equation to be
rewritten as
\begin{eqnarray*}
& &
\frac{\mathrm{d}}{\mathrm{d}\varepsilon}\Pi[\bu^\ast,\Bkappa_\varepsilon]\Big\vert_{\varepsilon=0}
=\\
& &\qquad\qquad\int\limits_{\Omega^\ast}\left[-\mathrm{Div}^\ast\left(\left(\bone-\bF^{\ast^\mathrm{T}}\right)\bP^\ast + \psi^\ast\bone + \BSigma^\ast\right)+\frac{\partial\psi^\ast}{\partial\bX^\ast} - \bff^\ast\right]\cdot\delta\Bkappa\mathrm{d}V^\ast\\
&
&\qquad\qquad+\int\limits_{\partial\Omega^\ast_t}\left[\left(\left(\bone-\bF^{\ast^\mathrm{T}}\right)\bP^\ast
+ \psi^\ast\bone + \BSigma^\ast\right)\cdot\bN^\ast -
\bar{\bt}^\ast\right]\cdot\delta\Bkappa\mathrm{d}A^\ast
\end{eqnarray*}

\noindent Enforcing equilibrium,
$\frac{\mathrm{d}}{\mathrm{d}\varepsilon}\Pi[\bu^\ast,\Bkappa_\varepsilon]\Big\vert_{\varepsilon=0}
= 0$, the localization principle and the arbitrariness of
$\delta\Bkappa$ give the following governing equations for
configurational change:

\begin{alignat}{2}
-\mathrm{Div}^\ast\left(\left(\bone-\bF^{\ast^\mathrm{T}}\right)\bP^\ast
+ \psi^\ast\bone +
\BSigma^\ast\right)+\frac{\partial\psi^\ast}{\partial\bX^\ast} -
\bff^\ast &= \bzero \quad\mathrm{in}\quad&&\Omega^\ast\\
\left(\left(\bone-\bF^{\ast^\mathrm{T}}\right)\bP^\ast +
\psi^\ast\bone + \BSigma^\ast\right)\bN^\ast - \bar{\bt}^\ast &=
\bzero \quad \mathrm{on}\quad&&\partial\Omega^\ast. \label{equil}
\end{alignat}

Using (\ref{quasistatic})$_1$ and (\ref{quasistatic})$_2$ these
equations are reducible to (\ref{remodequil}) and (\ref{remodbc}).

\bibliography{mybib}
\bibliographystyle{elsart-harv}







\end{document}